\documentclass[11pt]{article}
\usepackage{moriond,epsfig,amsmath}

\def\Journal#1#2#3#4{{#1} {\bf #2}, #3 (#4)}

\def\NPB{{\em Nucl.~Phys.}~B}
\def\PLB{{\em Phys.~Lett.}~B}

\def\PRD{{\em Phys.~Rev.}~D}

\def\epem{\ensuremath{\text{e}^+\text{e}^-}}
\def\as{\ensuremath{\alpha_{\scriptscriptstyle\text{S}}}}
\def\xe{\ensuremath{x_{\scriptscriptstyle \text{E}}}}
\def\qbar{\ensuremath{\bar{\text{q}}}}

\def\be{\begin{equation}}
\def\ee{\end{equation}}
\def\bea{\begin{eqnarray}}
\def\eea{\end{eqnarray}}

\begin{document}
\vspace*{4cm}
\title{LATEST QCD RESULTS FROM LEP}

\author{ M.T. FORD }

\address{Department of Physics and Astronomy, University of Manchester,\\
Oxford Road, Manchester, M13 9PL, U.K.}

\maketitle\abstracts{ We summarise the latest experimental QCD studies
based on data from LEP. Measurements of the quark
and gluon jet fragmentation functions are discussed, including a new
algorithm to infer the properties of unbiased gluon
jets. We describe a new test for destructive interference in the
radiation of soft gluons from a three-parton system. Finally, we
report the latest combined value of the strong coupling,
measured using event shape observables. }

\section{Introduction}

The LEP collider at CERN was used to study \epem\ annihilation at
centre-of-mass energies in the range $\sqrt{s}=91$--209~GeV, during
the years 1989--2000. Four multi-purpose detectors (ALEPH, DELPHI, L3
and~OPAL) were positioned at 90$^\circ$ intervals around the circular
accelerator. Events of the type
$\epem\to\text{Z}^0/\gamma\to\text{hadrons}$ were used extensively to
test QCD predictions, and to measure the colour factors and strong
coupling. In this article, we outline four of the most recent studies
performed by the LEP collaborations.

The text is organised as follows: in Sections~\ref{fragfun}
and~\ref{unbiased}, we report measurements of quark and gluon jet
properties, published by the OPAL Collaboration. Section~\ref{fragfun}
deals with the scaling of fragmentation functions over a range of
energies, for jet samples with a variety of flavour compositions,
while Section~\ref{unbiased} focuses on a new study of `unbiased'
gluon jets. In Section~\ref{coherence}, we discuss an innovative test
for the presence of destructive interference in the radiation of
particles from a three-jet system, by the DELPHI
Collaboration. Finally, in Section~\ref{alphas}, we report the progress
of a combined measurement of the strong coupling derived from event
shape observables, performed by the LEP QCD Working Group.

We emphasise that the results discussed here do not constitute an
exhaustive list of the recent and ongoing QCD studies by the LEP
collaborations. In particular, the results of pentaquark searches,
colour reconnection studies, and $\gamma\gamma$ physics will be
omitted due to time constraints.

\section{Scaling violations of quark and gluon jet fragmentation functions}
\label{fragfun}

The fragmentation function, $D_a^h(x,Q^2)$, is defined as the
probability that a parton $a$, which is produced at a short distance
of order $1/Q$, fragments into a hadron $h$ carrying a fraction $x$ of
the momentum of $a$. QCD predicts that the multiplicity of soft gluon
emission from a gluon source should be higher than that from a quark
source, due to the inequality of the colour factors $C_A$
and~$C_F$. We therefore expect softer fragmentation functions for gluon
jets. Furthermore, one can predict the dependences of the
fragmentation functions on the scale~$Q^2$, by means of the splitting
functions $P_{\text{q}\to\text{qg}}\sim C_F$
and~$P_{\text{g}\to\text{gg}}\sim C_A$: the scaling violations for
gluon jets are expected to be larger than those for quark jets.

In a recent study by the OPAL Collaboration,\cite{fragfun} the
fragmentation functions have been measured for a variety of quark and
gluon jet samples in
$\epem\to(\text{Z}^0/\gamma)\to\text{q}\bar{\text{q}}\text{(g)}$
interactions at \mbox{$\sqrt{s}=91.2$} and 183--209~GeV. There are two
experimental approaches to the identification of jets. Firstly, one
can use a jet-finding algorithm (the Durham algorithm in this case):
jets obtained by this method are {\em biased}, because their
properties depend on the choice of jet finder, and on its associated
parameters. Alternatively, jets may be defined as inclusive
hemispheres of particles in a back-to-back q$\bar{\text{q}}$ or gg
system: these {\em unbiased} jets correspond to the definitions
commonly used in theoretical calculations. The new OPAL measurements
include seven types of fragmentation functions: the udsc, b, gluon and
flavour-inclusive fragmentation functions for biased jets, and the
udsc, b and flavour-inclusive quark fragmentation functions for
unbiased jets. A previous study of unbiased gluon jets has been
performed with rare OPAL events of the type
$\epem\to\text{q}\bar{\text{q}}\text{g}_\text{incl}$, in which the
gluon `$\text{g}_\text{incl}$' is identified as the hemisphere
opposite to a hemisphere containing two tagged b~quark jets which are
almost collinear.\cite{oldunbiasedgluon} The new results presented
here for biased gluon jets will be compared with the older results for
unbiased jets. An alternative approach to unbiased gluon jets will be discussed in the next
section.

Three methods have been used to distinguish between quark and gluon
jets, and between different flavours of quark jets. In the {\em
b-tag~method}, a neural network based algorithm is used to select
b~quark jets in three-jet events. The untagged jet in an event
containing two b-tagged jets (or the lowest-energy untagged jet in an
event with one b-tag) is selected as a biased gluon jet. In events
with no b-tagged jets, all three jets are selected as biased udsc
quark jets; a correction is applied to remove gluon jets from the
sample. In the {\em energy-ordering method}, the jets 1, 2 and~3 of a
three-jet event are ordered such at $E_1>E_2>E_3$. Jet number~2 is
assigned to the biased flavour-inclusive quark jet sample, while jet
number~3 is assigned to the biased gluon jet sample.  Finally, in the
{\em hemisphere method}, a b-tagging algorithm is applied to an
inclusive sample of hadronic events. Each event contains two
hemispheres, which are regarded as unbiased udsc or b quark jets.  In
all three methods, an unfolding procedure is employed to correct for
impurities in the selection.

For experimental purposes, we define the scale $Q$ for a biased jet of
energy $E_\text{jet}$ to be $Q_\text{jet}=E_\text{jet}\sin(\theta/2)$,
where $\theta$ is the angle to the closest jet. In the case of
unbiased jets, we take $Q=\sqrt{s}/2$. The momentum fraction~$x$ is
replaced by the quantity $\xe=E_\text{h}/E_\text{jet}$, where
$E_\text{h}$ is the energy of the hadron and $E_\text{jet}$ is the
energy of the jet to which it is assigned.

Next-to-leading order QCD predictions for the fragmentation functions
have been calculated by three groups: Kniehl, Kramer and P\"otter
(KKP),\cite{fragfun_kkp} Kretzer (Kr),\cite{fragfun_kr} and Bourhis,
Fontannaz, Guillet and Werlen (BFGW).\cite{fragfun_bfgw} The
predictions correspond to unbiased jets, and are derived using the
DGLAP evolution equations, from a set of measured fragmentation
functions at a fixed input scale.

\begin{figure}[p]
\begin{center}
\psfig{figure=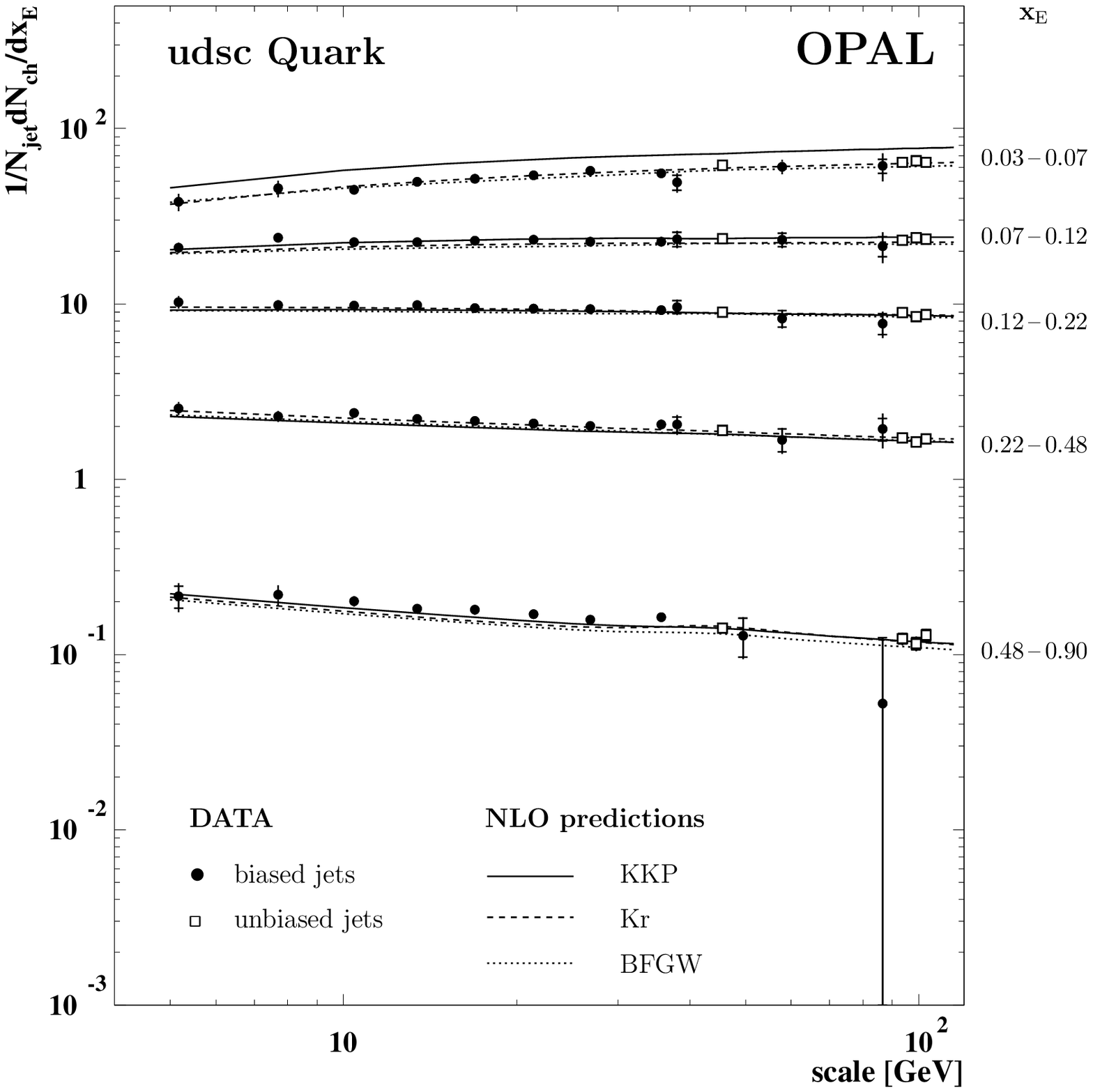,height=4.3in}\\[0.2in]
\psfig{figure=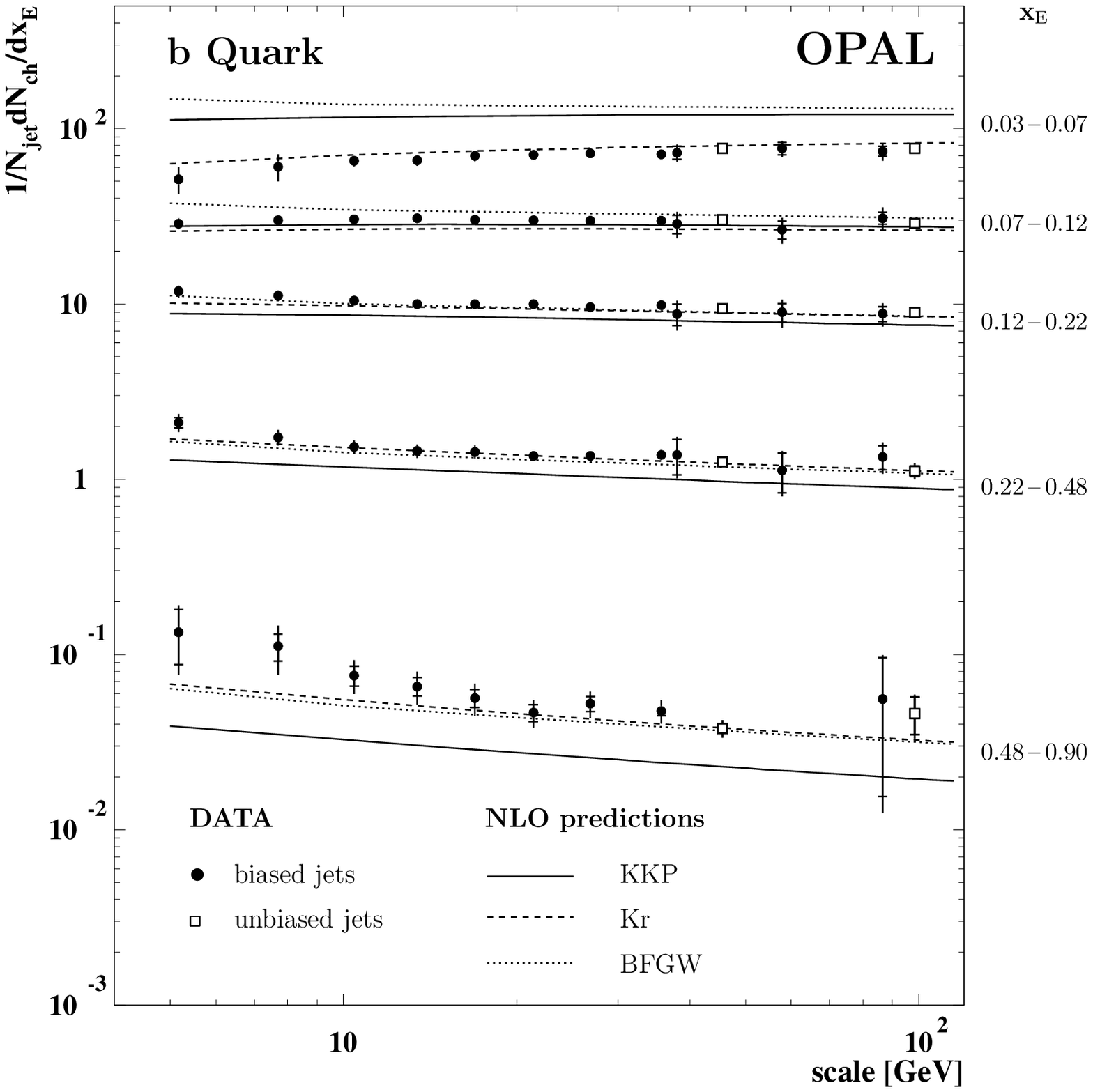,height=4.3in}
\end{center}
\caption{Scale dependence of the udsc and b jet fragmentation
functions in different bins of \xe. The `scale' denotes $Q_\text{jet}$
for the biased jets and $\sqrt{s}/2$ for the unbiased jets. The inner
and outer error bars indicate statistical and total uncertainties
respectively. The data are compared to NLO predictions by
KKP,\protect\cite{fragfun_kkp} Kr \protect\cite{fragfun_kr} and
BFGW.\protect\cite{fragfun_bfgw}}
\label{fig:fragfun_udscb_q}
\end{figure}

\begin{figure}[t]
\begin{center}
\psfig{figure=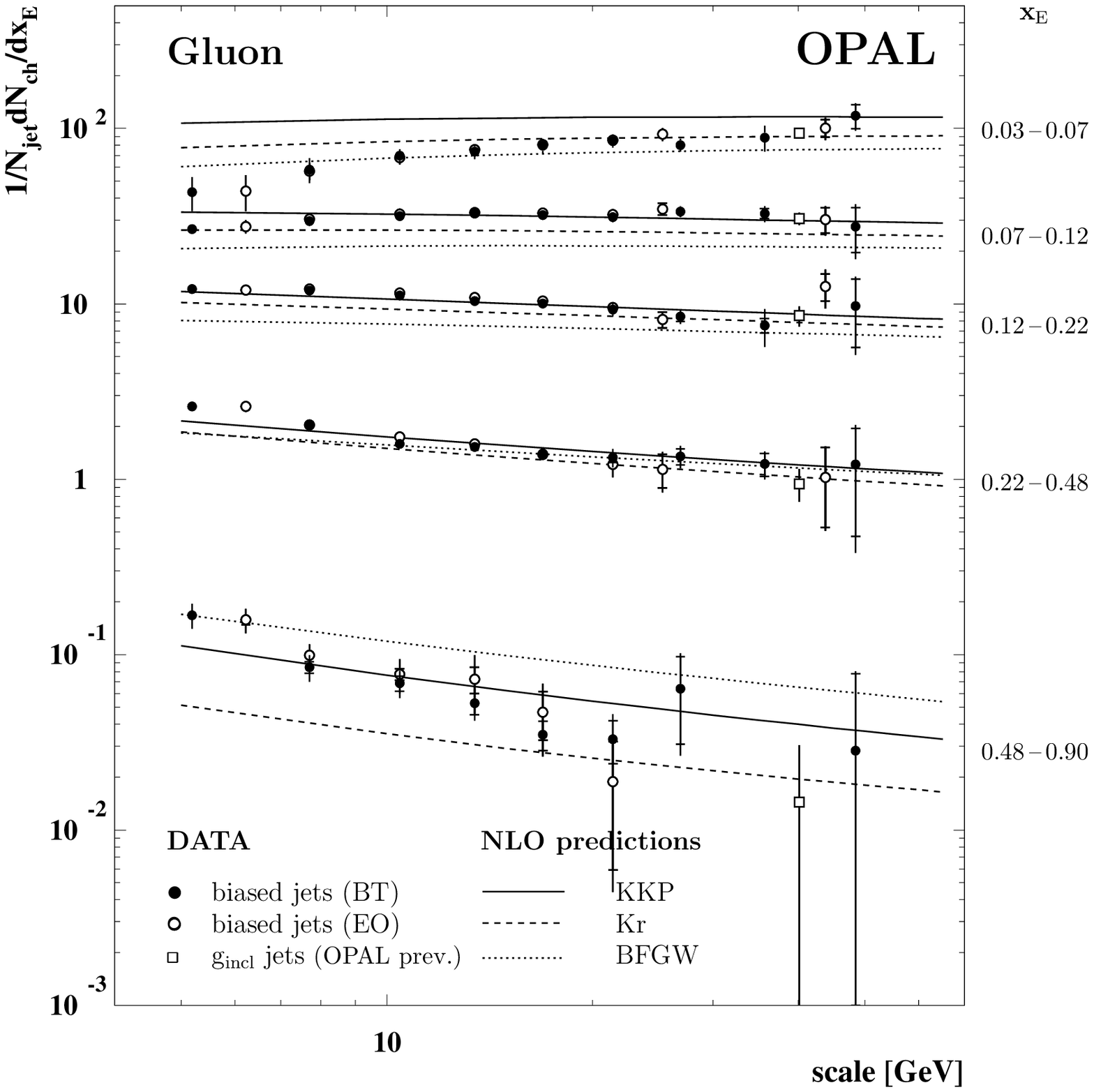,height=4.3in}
\end{center}
\caption{Scale dependence of the gluon jet fragmentation functions in
different bins of \xe. The `scale' denotes $Q_\text{jet}$ for the
biased jets and $E_\text{jet}$ for the previously published results
using unbiased `$\text{g}_\text{incl}$' jets. The results for biased
jets are shown separately for the the b-tag~(BT) and
energy-ordering~(EO) selection methods. The inner and outer error bars
indicate statistical and total uncertainties respectively.}
\label{fig:fragfun_g}
\end{figure}

The measured scale dependences for udsc, b and gluon fragmentation
functions are shown in Figures~\ref{fig:fragfun_udscb_q}
and~\ref{fig:fragfun_g}, for different ranges of~\xe. For the udsc
quark fragmentation functions, good agreement is found with the NLO
predictions, except in the lowest and highest regions of \xe. The
agreement is poorer in the case of b~quark and gluon jets; however,
the gluon jet fragmentation functions do exhibit stronger scaling
violations than the quark jets, as expected. Good agreement is found
between the biased and unbiased jet samples, suggesting that
$Q_\text{jet}$ is an appropriate scale in events with a three-jet
topology. The results obtained using the b-tag and energy-ordering
methods are also consistent with one another, and with previous
results from DELPHI and OPAL.

\section{Studies of unbiased gluon jets using the jet boost algorithm}
\label{unbiased}

\begin{figure}[t]
\begin{center}
\includegraphics[bb=60 567 548 755, clip=true, height=1.5in]{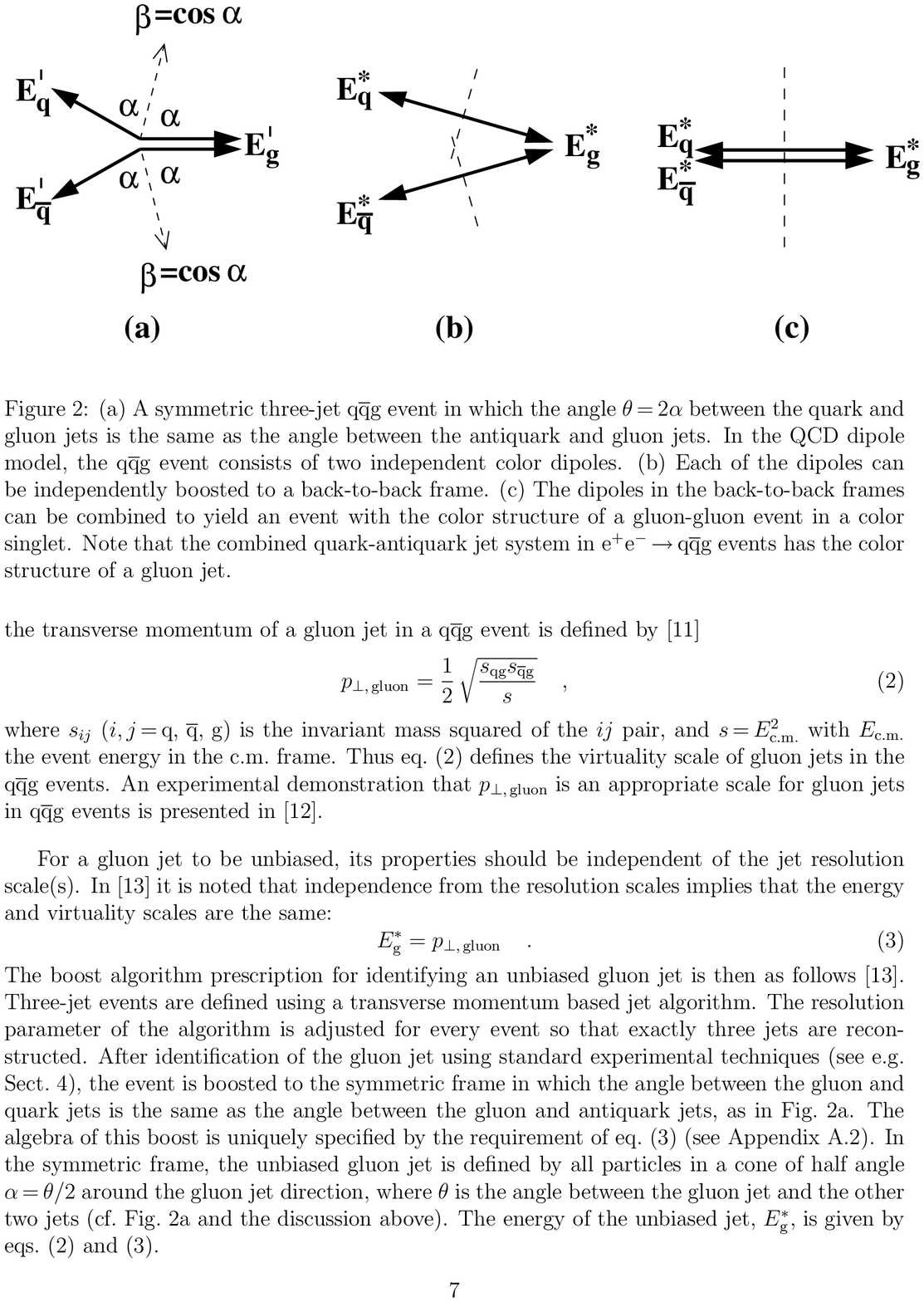}
\vspace{-0.3cm}
\end{center}
\caption{The jet boost algorithm}
\label{fig:jetboost}
\end{figure}

As we have mentioned in Section~\ref{fragfun}, a distinction exists
between the biased jets obtained from experimental data using a
jet-finding algorithm, and the unbiased jets used in theoretical
calculations. An unbiased gluon jet would correspond to one hemisphere
of a back-to-back gg system, which is not seen in \epem\
annihilation. Instead, rare events of the type
$\epem\to\text{q}\bar{\text{q}}\text{g}_\text{incl}$ have been used,
in which the q and $\bar{\text{q}}$ jets are approximately collinear,
leaving an unbiased `g$_\text{incl}$' jet in the opposite
hemisphere.\cite{oldunbiasedgluon} Another OPAL study has used a more
indirect method, whereby the results obtained from two-jet q\qbar\
events are subtracted from those in q\qbar g
events.\cite{gluon_subtract} Unbiased gluon jets have also been
obtained from radiative $\Upsilon\to\gamma\text{gg}$ decays at
CLEO.\cite{cleo} Recently, however, a new approach known as the {\em
jet boost algorithm} has been considered in \epem\
annihilation.\cite{boost_theory} In this method, a q\qbar g system is
decomposed into two independent qg and \qbar g dipoles. The dipoles
are boosted into a symmetric frame, such that the angle $2\alpha$
between the q and g is the same as that between the \qbar\ and~g, as
shown in Figure~\ref{fig:jetboost}(a). Further Lorentz
boosts~$\beta=\cos\alpha$ are then applied independently to the two
dipoles, such that they are each back-to-back, as in
Figure~\ref{fig:jetboost}(b). Finally the two dipoles in their
different frames can be recombined, yielding an event with the
colour-structure of a gg system in a colour singlet state, as shown in
Figure~\ref{fig:jetboost}(c). The hemisphere containing the gluon in
this event corresponds to the theoretical definition of an unbiased
gluon jet. Unlike the unbiased quark jets discussed in
Section~\ref{fragfun}, the energies $E_\text{g}^*$ of these gluon jets
are not fixed by the \epem\ centre-of-mass energy.

The jet boost algorithm has recently been studied by the OPAL
Collaboration, using $\epem\to\text{q}\qbar\text{g}$ events at the
Z$^0$ resonance.\cite{unbiasedgluon} Gluon jets are selected using a
combination of b-tagging and energy-ordering; cuts then are imposed on
the softness and collinearity of the quark and gluon jets. Using the
HERWIG Monte Carlo event generator, it has been established that the
unbiased gluon jets obtained using the jet boost algorithm should have
properties consistent with a true back-to-back gg pair. It has also
been shown that the properties are essentially independent of the
jet-finder used to identify the q, \qbar\ and g jets, as expected for
an unbiased jet.

\begin{figure}[t]
\begin{center}
\vspace{0.2cm}
\includegraphics[height=2.5in]{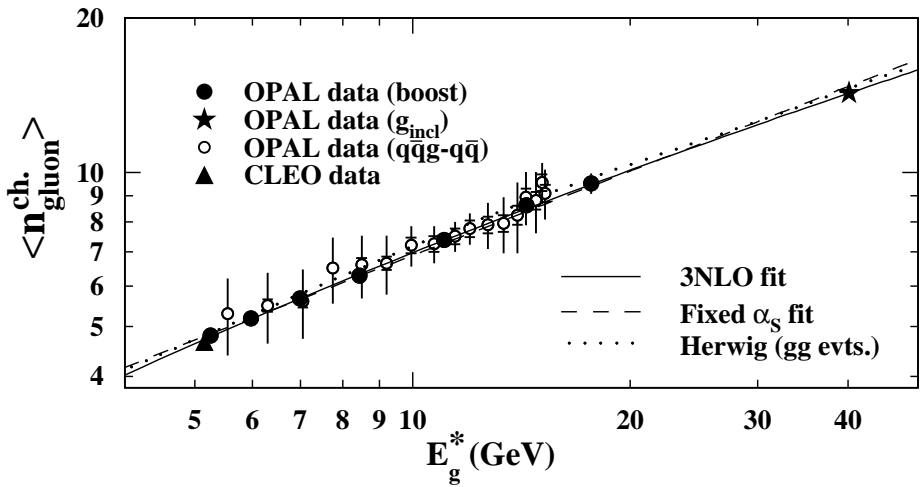}
\vspace{-0.3cm}
\end{center}
\caption{The mean charged particle multiplicity of unbiased gluon
jets, $\left\langle n_\text{gluon}^\text{ch.}\right\rangle$, as a
function of the jet energy~$E_\text{g}^*$.}
\label{fig:nch}
\end{figure}

The mean charged multiplicity of the unbiased gluon jets, $\langle
n_\text{gluon}^\text{ch.}\rangle$, is shown in Figure~\ref{fig:nch},
as a function of the jet energy~$E_\text{g}^*$. The results are found
to be consistent with previous OPAL measurements based on the
`g$_\text{incl}$' and `subtraction'
methods,\cite{oldunbiasedgluon,gluon_subtract} and with the HERWIG
prediction for genuine gg pairs. QCD evolution fits are also performed
for two calculations: one based on next-to-next-to-next-to-leading
order (3NLO) perturbation theory for a scale-dependent \as, and the
other based on an exact solution for a fixed \as. Two free parameters
are used in each case, and no hadronisation corrections are
applied. The QCD scale parameter $\Lambda$ is hence found to be
$\Lambda=0.296\pm 0.037\,$(total)~GeV, compared to the corresponding
quark jet result $\Lambda=0.190\pm
0.032\,$(stat.)~GeV.\cite{gluon_subtract} Similar analyses have been
performed for the first two non-trivial factorial moments of the
charged particle multiplicity distribution, $F_\text{2,gluon}$ and
$F_\text{3,gluon}$.\cite{unbiasedgluon}

Fragmentation functions have also been obtained for the unbiased gluon
jets, at energies $E_\text{g}^*$=14.24 and 17.72~GeV. Since the OPAL
Collaboration has previously measured the fragmentation functions for
unbiased gluon jets at 40.1~GeV,\cite{oldunbiasedgluon} and for
unbiased quark jets at 45.6~GeV,\cite{oldunbiasedquark} the DGLAP
evolution equations can be used to construct a QCD prediction for the
new measurements at lower energies. Such predictions are dependent on
the strong coupling~\as, which can therefore be extracted from a
one-parameter fit. The fits are in good agreement with
the measured fragmentation functions, yielding
\mbox{$\as(M_\text{Z})=0.128\pm 0.008\,\text{(stat.)}\pm
0.015\,\text{(syst.)}$}. \mbox{Although} not competitive with other
measurements of the strong coupling, this result is compatible with
the world average, and has provided a unique consistency test of QCD.

\section{Coherent soft particle production in \boldmath Z$^0$ decays into three jets}
\label{coherence}

Interference effects are fundamental to all quantum mechanical
theories, including the gauge theories of the Standard Model. Evidence
for coherence effects in QCD comes, for example, from the
``hump-backed plateau'' in the logarithmic scaled momentum spectrum of
hadrons, due to suppression of low energy particle production, and
from the string effect in three-jet \epem\ annihilation
events.\cite{coherencereview} However, there are arguments against the
conclusiveness of \mbox{this evidence.\cite{boudinov}}

In a new study by the DELPHI Collaboration,\cite{delphinote} a direct
test is made for the presence of destructive interference in the
radiation of soft gluons from a three-jet system. When a quark emits a
hard gluon at a small opening angle, the quark-gluon system may behave
on large distance scales as a single entity. A soft gluon radiated
perpendicular to the q--g pair will not be able to resolve the
individual colour charges of the quark and gluon; it should therefore
be regarded a coherent emission from the parton ensemble. The leading
order cross section, d$\sigma_3$, has been calculated for coherent
soft gluon emissions perpendicular to the plane of a q\qbar g system:~\cite{partonperp}
\begin{equation}
\text{d}\sigma_3=\text{d}\sigma_2\cdot
\frac{C_A}{4C_F}
\left[\widehat{\text{qg}}+\widehat{\qbar\text{g}}-\frac{1}{N_\text{c}^2}\widehat{\text{q}\qbar}\right]\;\;\;.
\label{eq:coherence}
\end{equation}
Here d$\sigma_2$ represents the corresponding cross section for soft
gluon emissions perpendicular to the axis of a two-jet (q\qbar) event,
and the antenna function $\widehat{ij}$ for a pair of partons $(i,j)$
with an opening angle $\theta_{ij}$ is defined by
$\widehat{ij}=1-\cos\theta_{ij}$.
The last term of Equation~(\ref{eq:coherence}), inversely proportional
to the square of the number of colours $N_\text{c}=3$, is due to destructive
interference. By measuring the ratio
$\text{d}\sigma_3/\text{d}\sigma_2$ in samples of two- and three-jet events,
one should be able to verify this
$(1/N_\text{c}^2)\,\widehat{\text{q}\qbar}$ interference term, and to
determine the colour factor ratio $C_A/C_F$ based on the leading order
expression given above.

\begin{figure}[t]
\begin{center}
\vspace{0.2cm} \includegraphics[height=5in,clip=true]{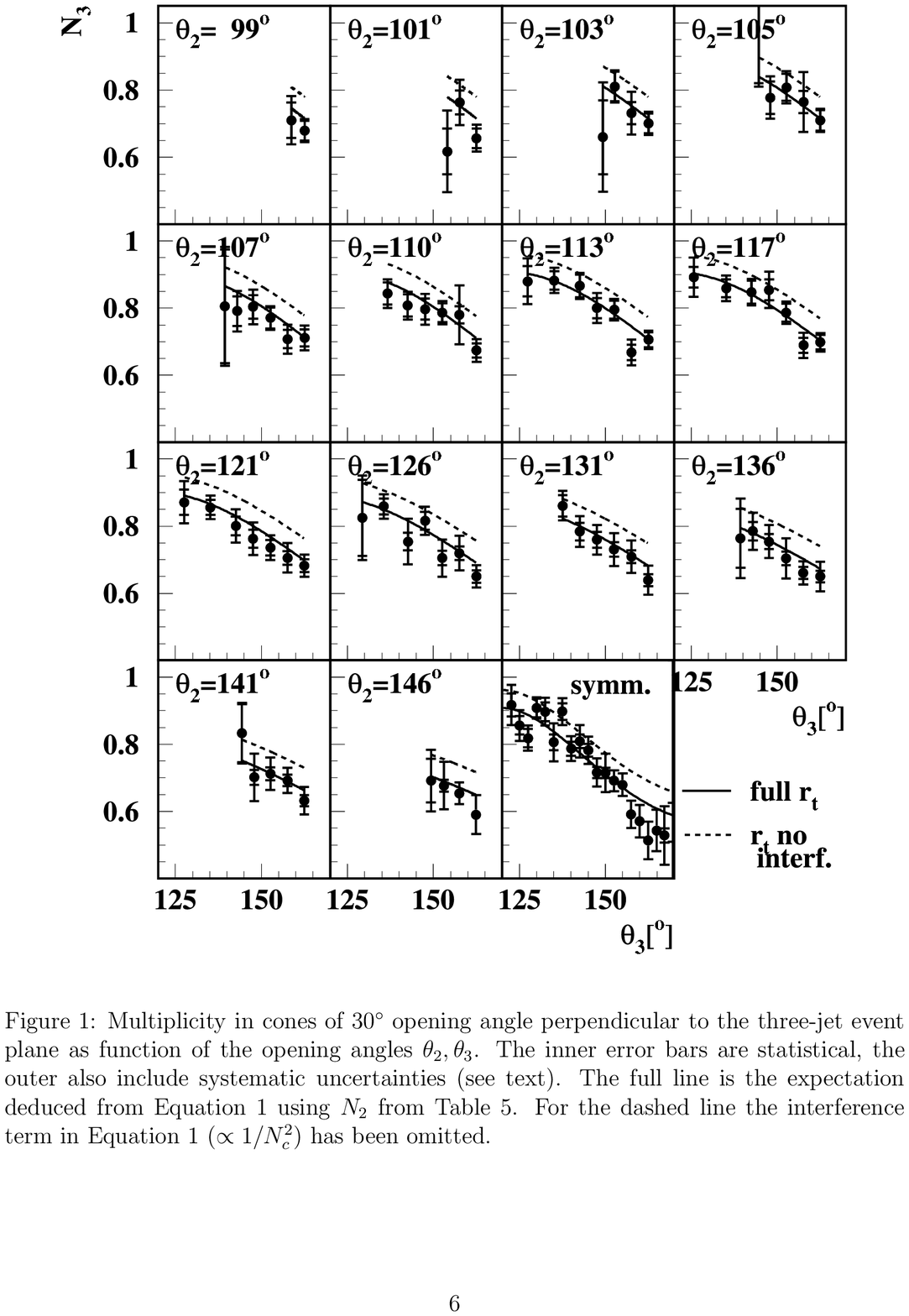}
\vspace{-0.3cm}
\end{center}
\caption{Particle multiplicity in cones of 30$^\circ$ opening angle
perpendicular to the three-jet event plane, as functions of the opening
angles $\theta_2,\theta_3$. The inner error bars are statistical,
while the outer bars include both statistical and systematic
uncertainties.}
\label{coherenceplot}
\end{figure}

Hadronic Z$^0$ decays were recorded by the DELPHI detector in \epem\
annihilation at $\sqrt{s}=91$~GeV. The angular ordered Durham
algorithm~\cite{aod} was used, with a fixed resolution parameter
$y_\text{cut}=0.015$, to determine the number of jets in each event.
For each three-jet event, the charged particle multiplicity $N_3$ was
measured in a 30$^\circ$ cone oriented perpendicular to both sides of
the event plane. The corresponding multiplicity $N_2$ was also
measured in two-jet events, for a cone perpendicular to the event
axis; the azimuthal angle of the cone was chosen randomly in this
case. In Figure~\ref{coherenceplot}, we show the three-jet cone
multiplicity~$N_3$ as a function of the inter-jet angles $\theta_2$
and $\theta_3$, where $\theta_i$ is the angle between the two jets
opposite to jet~$i$, and the jets are energy-ordered such that
$E_1<E_2<E_3$. From the measured value of $N_2$, a prediction can be
calculated for $N_3$ using Equation~(\ref{eq:coherence}): these
predictions are shown in Figure~\ref{coherenceplot}, with and without
the destructive interference term. The expression without the
$(1/N_\text{c}^2)\,\widehat{\text{q}\qbar}$ term is incompatible with
the data, while the fully coherent prediction is in good agreement. If
the ratio of colour factors, $C_A/C_F$, is fitted to the data, one
obtains the value $2.211\pm 0.014\,\text{(stat.)}\pm
0.053\,\text{(syst.)}$ with $\chi^2/\text{ndf}=1.3$. Although this
result is valid only at leading order, and does not include an
estimate of the theoretical uncertainty, it is in astonishingly good
agreement with the expectation $C_A/C_F=2.25$. The leading order QCD
prediction for the soft gluon multiplicity, including destructive
interference, has therefore been verified convincingly by the data.

\section{A combined measurement of \boldmath\as\ using event shape observables}
\label{alphas}

Event shape observables have been used extensively to test QCD
predictions, and to measure the strong coupling \as, in both \epem\
annihilation and deep inelastic scattering.\cite{dasgupta} The LEP
collaborations have published measurements of \as\ for events in the
energy range \mbox{$\sqrt{s}=91$--209~GeV,\cite{thesis}} using the
following six event shape observables: thrust~($T$), heavy jet
mass~($M_\text{H}$), $C$-parameter, total jet
broadening~($B_\text{T}$), wide jet broadening~($B_\text{W}$), and
the Durham two-to-three jet resolution parameter~($y_3$). For each
distribution, an $\mathcal{O}(\as^2)$ perturbative prediction is
matched with an NLLA resummation and corrected for hadronisation
effects using the PYTHIA Monte Carlo event generator. The QCD
predictions are then fitted to the LEP data, with \as\ as a free
parameter. The LEP QCD Working Group has combined these into a single
preliminary\hspace{0.08cm}\footnote{A publication is expected later in 2004, when
measurements provided by the collaborations have become final.}
measurement of \as\ at the Z$^0$ mass scale, with particular attention
to estimation of theoretical uncertainties~\cite{lepqcdtheory} and to
the treatment of correlations.\cite{thesis} The result is
\[
\as(M_\text{Z})=0.1202\pm 0.0003~\text{(stat.)}\pm 0.0009~\text{(expt.)}
\pm 0.0013~\text{(hadr.)}\pm 0.0047~\text{(theo.)}\;\;\;,
\]
where the three systematic uncertainties are due to experimental
effects~(`expt.'), Monte Carlo hadronisation corrections~(`hadr.')~and
higher-order terms in the perturbative QCD predictions~(`theo.'). Our
measurement is in good agreement with the current world average.\cite{pdg}

\begin{figure}[t]
\begin{center}
\includegraphics[height=3.8in, clip=true]{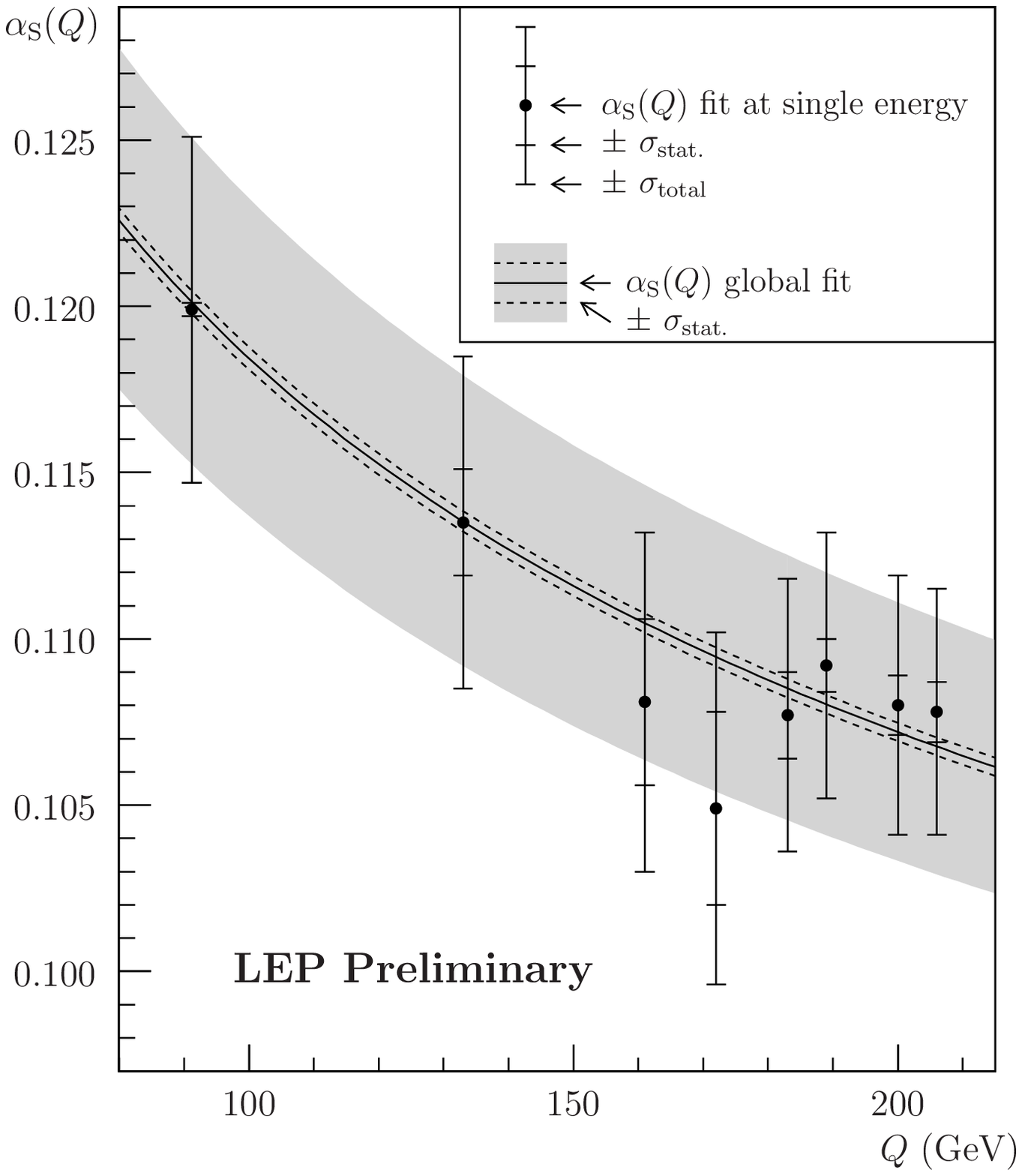}
\hspace{0.25cm}\includegraphics[height=3.8in, clip=true]{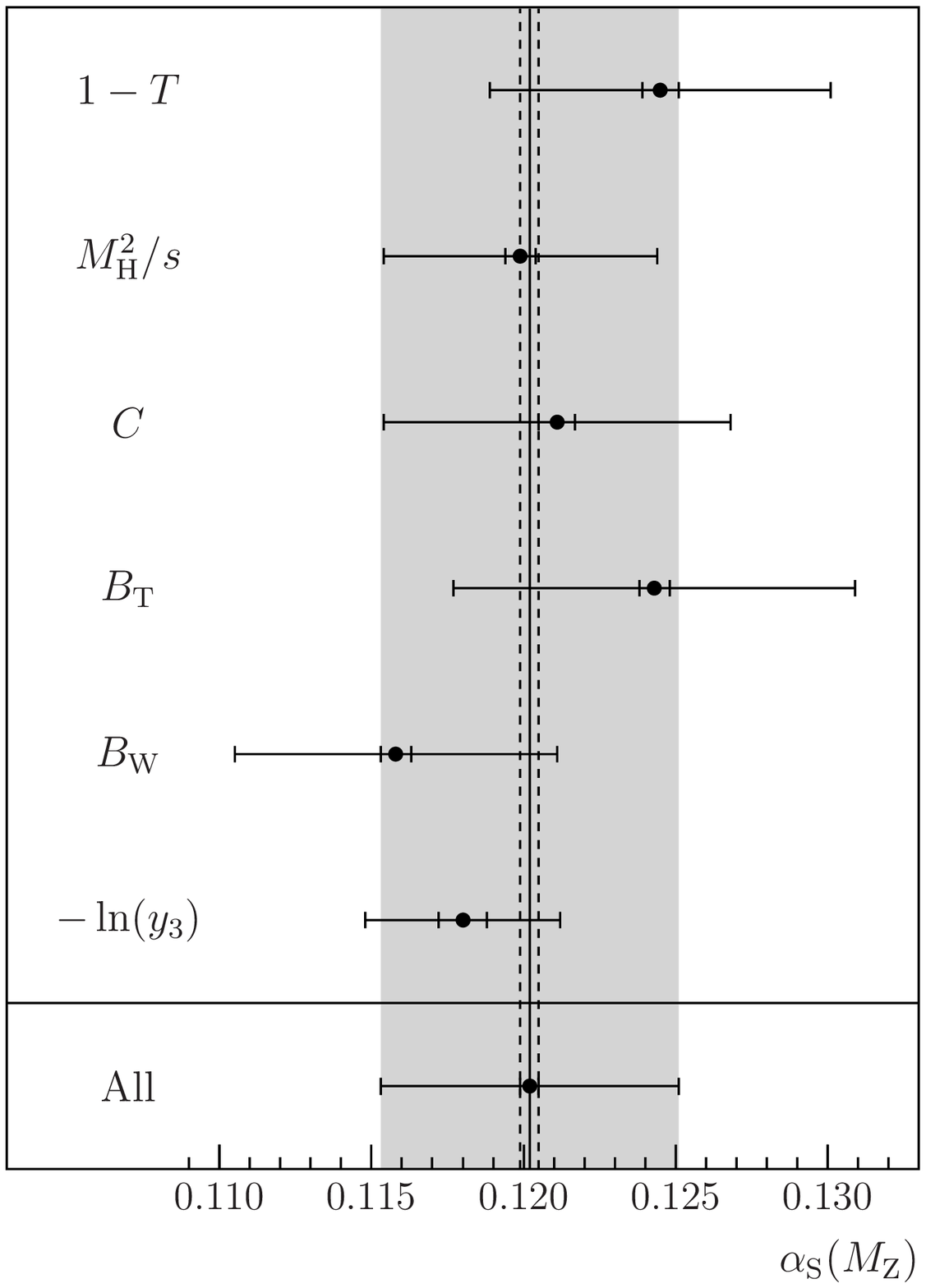}
\vspace{-0.3cm}
\end{center}
\caption{{\em Left figure:} Measurements of the strong coupling
$\as(Q)$ using event shape observables at LEP. Each point is a
combination of all available measurements at a single energy; the
inner error bar represents the combined statistical uncertainty, while
the outer bar is the total uncertainty. The curve indicates the
running of $\as(Q)$ predicted by QCD at three-loop order, based on our
combined measurement of \as\ at the Z$^0$ mass scale,
$\as(M_\text{Z})$. The statistical and total uncertainties in
$\as(M_\text{Z})$ are indicated by the dotted curve and grey band
respectively. {\em Right~figure:}~Combined measurements of
$\as(M_\text{Z})$, using individual event shape observables.\vspace{-0.2cm}}
\label{asrunning}
\end{figure}

\end{document}